\documentclass[twocolumn]{aastex631}

\usepackage{graphicx}
\usepackage{amsmath,amssymb}
\usepackage{float}
\usepackage{xcolor}
\usepackage{hyperref}

\graphicspath{{./}{figs}}

\newcommand{\nubh}{$\nu\texttt{bhlight}$}

\received{\today}
\submitjournal{\apj}

\shorttitle{Short versus long accretion disk nucleosynthesis}
\shortauthors{Sprouse et al.}

\begin{document}

\title{Emergent nucleosynthesis from a 1.2 second long simulation of a black-hole accretion disk}

\correspondingauthor{M. R. Mumpower}
\email{mumpower@lanl.gov}

\author[0000-0002-4375-4369]{Trevor M.\ Sprouse}
\affiliation{Theoretical Division, Los Alamos National Laboratory, Los Alamos, NM 87545, USA}
\affiliation{Center for Theoretical Astrophysics, Los Alamos National Laboratory, Los Alamos, NM 87545, USA}

\author[0000-0003-0031-1397]{Kelsey A. Lund}
\affiliation{Department of Physics, North Carolina State University, Raleigh, NC 27695, USA}
\affiliation{Theoretical Division, Los Alamos National Laboratory, Los Alamos, NM 87545, USA}
\affiliation{Center for Nonlinear Studies, Los Alamos National Laboratory, Los Alamos, NM 87545, USA}

\author[0000-0001-6432-7860]{Jonah M.\ Miller}
\affiliation{Computational Division, Los Alamos National Laboratory, Los Alamos, NM 87545, USA}
\affiliation{Center for Theoretical Astrophysics, Los Alamos National Laboratory, Los Alamos, NM 87545, USA}

\author[0000-0001-6811-6657]{Gail C. McLaughlin}
\affiliation{Department of Physics, North Carolina State University, Raleigh, NC 27695, USA}

\author[0000-0002-9950-9688]{Matthew R.\ Mumpower}
\affiliation{Theoretical Division, Los Alamos National Laboratory, Los Alamos, NM 87545, USA}
\affiliation{Center for Theoretical Astrophysics, Los Alamos National Laboratory, Los Alamos, NM 87545, USA}

\begin{abstract}
We simulate a black-hole accretion disk system with full-transport general relativistic neutrino radiation magnetohydrodynamics (GR$\nu$RMHD) for 1.2 seconds. 
This system is likely to form after the merger of two compact objects and is thought to be a robust site of $r$-process nucleosynthesis. 
We consider the case of a black-hole accretion disk arising from the merger of two neutron stars. 
Our simulation time coincides with the nucleosynthesis timescale of the $r$ process ($\sim$ 1 second). 
Because these simulations are time consuming, it is common practice to run for `short' duration of approximately 0.1 to 0.3 seconds. 
We analyze the nucleosynthetic outflow from this system and compare the results between stopping at 0.12 and 1.2 seconds respectively. 
We find that the addition of mass ejected in the longer simulation as well as more favorable thermodynamic conditions from emergent viscous ejecta greatly impacts the nucleosynthetic outcome. 
We quantify the error in nucleosynthetic outcomes between short and long cuts.

\end{abstract}



\section{Introduction}
\label{sec:intro}

The outflows of a black-hole accretion disks are promising sites for the synthesis of the heavy elements via the rapid neutron capture ($r$ process) \citep{Freiburghaus1999, Siegel2017, SiegelMetzger2018, Nouri2018, Miller2019a, Miller2019b, FoucartMerger2020, Kullmann2021,Foucart2021,Fahlman2022,Just2022}. 
Such environments may form after the merger of compact objects and potentially offer unique signatures of heavy element formation \citep{Metzger2010, Zhu2018, Korobkin2020, Zhu2021, Lund2023}.

Accretion disks have been modeled in increasing detail for many years with notable works from \citet{Pringle1972, Ruffert1996, Popham1999, MacFadyen1999, Shibata2000}.

The behavior of accretion disks is sensitive to a number of physical effects including post-merger magnetic field configurations \citep{Rudiger2002, Christie2019}, the nuclear equation of state \citep{Steiner2013}, and neutrino physics \citep{Mclaughlin2005, Surman2008}. 
In neutron star mergers, disk ejecta may be accompanied by dynamical ejecta \citep{Dietrich2017, Radice2018} that is also sensitive to neutrino physics \citep{Foucart2023CollapsingNS}. 
Accretion disks from the merger of a neutron-star black-hole binary are also found to be favorable sites of the $r$ process \citep{Siegel2017, De2021, Murguia-Berthier2021, Curtis2023}.

Long term evolution of accretion disks is consequential for electromagnetic counterparts \citep{Fernandez2019,Christie2019} as well as the nucleosynthesis that ensues in the aftermath of these cataclysmic events. 
Recently significant effort has been devoted to simulations that capture the long-lived remnant \citep{KotaSecondsLong2022,KotaSecondsLong2023}. 
To our knowledge, however, no late-time models to date perform detailed radiation transport and nucleosynthesis calculations. 
Previous work \citep{Miller2019b, Miller2020} indicates that at early times, higher-fidelity transport is required to accurately capture the electron fraction of the outflow and thus the nucleosynthetic yields. 
To date it is unclear if this result translates to late times during active nucleosynthesis. 

In this work, we help resolve this uncertainty. We model a black-hole accretion disk system that may arise after the merger of two neutron stars and evolve it for 1.2 seconds. 
This duration of time is long enough to explore active nucleosynthesis in the $r$ process. 
We analyze mass ejection, entropy, and electron fraction which all have a strong influence on the nucleosynthetic outcomes. 
To analyze the error in present model calculations arising from computational limitations, we compare these results to the same simulation stopped at 0.12 seconds. 
We end with a discussion of the uncertainty that arises in simulated nucleosynthesis yields when using short-duration simulations. 

\section{Simulating nucleosynthesis}

\subsection{Simulation details}

We extend the full transport general relativistic neutrino radiation magnetohydrodynamics (GR$\nu$RMHD) simulation of a black-hole accretion disk-wind system performed in \citep{Miller2019b} using the \nubh{} code \citep{Miller2019a, Miller2019b, Miller2020} to a full 1.2 seconds. This calculation took approximately 7 months of walltime.

The original model, which we extend, uses a stationary \citet{KerrBH} black hole
spacetime for a black hole of mass $M_{\rm BH} = 2.58 M_\odot$
and dimensionless spin $a=0.69$. The initial conditions are
a torus in hydrostatic equilibrium \citep{FishboneMoncrief} of
constant specific angular momentum, constant entropy of
$s = 4 k_b/$baryon, constant electron fraction $Y_{\rm e}=0.1$, and total
mass of $M_{\rm d} = 0.12 M_\odot$. Our torus starts with a single poloidal magnetic
field loop with a minimum ratio of gas to magnetic pressure, $\beta$, of 100.

We solve the equations of general relativistic ideal
magnetohydrodynamics, closed with the SFHo EOS, described in
\citet{SFHoEOS} and tabulated in \citet{stellarcollapsetables}.
Neutrinos are evolved with a Monte Carlo method and can interact with matter via emission, absorption, or
scattering. 
For emission and absorption, 
we use the charged and
neutral current interactions as tabulated in \citet{fornax} and
summarized in \citet{BurrowsNeutrinos}. Neutrino scattering is implemented
as described in \citet{Miller2019a}. The Monte Carlo and Finite Volume methods are coupled via first-order operator splitting.

We use a radially logarithmic, quasi-spherical grid in horizon penetrating coordinates
with $N_r\times N_\theta \times N_\phi = 192\times 168\times 66$ grid
points with approximately $3.8\times 10^7$ Monte Carlo packets. For details on the resolution requirements of the model, and why we chose this resolution, see \citet{Miller2019b}. After about 400 ms of runtime, the neutrino opacity in the disk is sufficiently low that neutrinos are essentially free-streaming. At this point, we turn off transport and switch to an optically thin cooling prescription. Essentially Monte Carlo particles are emitted at the proper rate but are then immediately deleted and not transported or absorbed. 

Although our code is Eulerian, we track approximately $1.5\times 10^6$ Lagrangian fluid packets, or ``tracer particles.'' 
Each tracer particle is assigned a mass, representing the statistical weight of the particle. 
Following \citet{Bovard2017}, we initialize tracer particles uniformly distributed in the volume containing a non-trivial density of gas at the initial time. 
At each time-step tracer particles are advected with the fluid flow via the equation 
\begin{equation}
    \label{eq:tracer:eqn}
    \frac{\partial x^i}{\partial t} = \frac{u^i}{u^0} = \alpha v^i - \beta^i
\end{equation}
for fluid four-velocity $u^\mu$, three-velocity $v^i$, lapse $\alpha$, and shift $\beta^i$. Latin indices range from 1 to 3 and represent spatial directions. Greek indices range from 0 to 3 and represent space and time.
Fluid and microphysical data, such as fluid density and temperature, electron fraction, and neutrino reaction rates are interpolated to tracer positions and recorded per tracer. 

\subsection{Engine Physics}

\begin{figure}[t]
  \centering
  \includegraphics[width=\columnwidth]{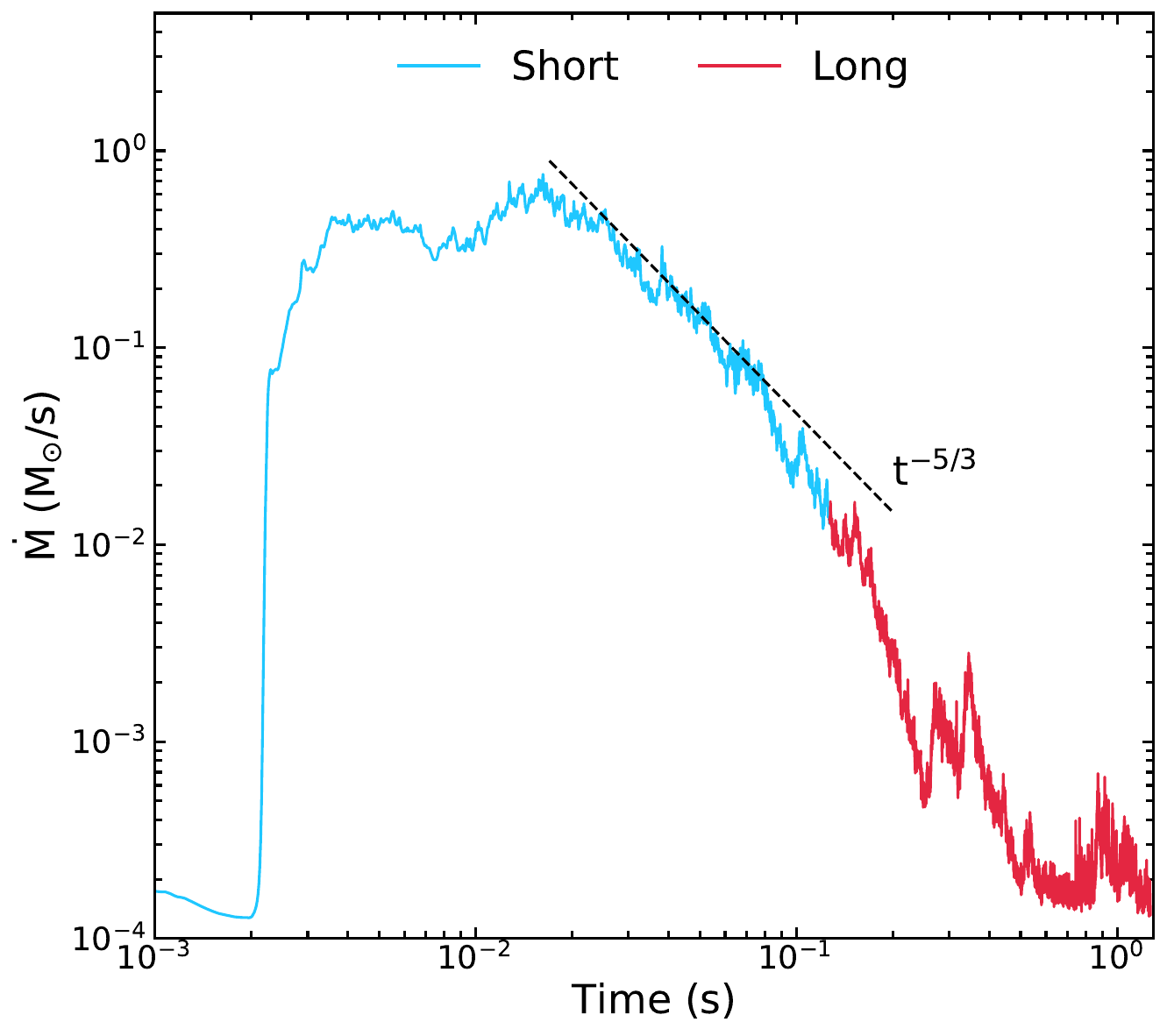}
  \caption{Accretion rate of the disk over the lifetime of the calculation. The blue line segment shows the accretion rate over the duration of the ``short" cut; the red line segment shows the extended part of the calculation, referred to as the ``long" cut.}
\label{fig:mdot}
\end{figure}

As the system evolves, the magneto-rotational instability
\citep[MRI,][]{VelikovMRI,BalbusHawley91} self-consistently drives the disk to a
turbulent state, which provides the turbulent viscosity necessary for
the disk to accrete \citep{ShakuraSunyaev1973}. 
This mechanism drives a long-lived accretion flow, which starts as powerful as $>1 M_\odot/s$ but sweeps down in accretion rate as the disk expands and cools. 
Figure \ref{fig:mdot} shows this behavior. 
Analytic models of the turbulent viscosity predict the accretion rate follows a $t^{-5/3}$ power law before eventually eventually transitioning to exponential decay \citep{Tanaka2011,Dolence2011}. 
We include a $t^{-5/3}$ line to guide the eye. 
Material undergoing $r$-process nucleosynthesis is ejected primarily during the downward sloping phase of this curve, after approximately $2 \times 10^{-2}$ seconds. 

\begin{figure}[t]
  \centering
  \includegraphics[width=\columnwidth]{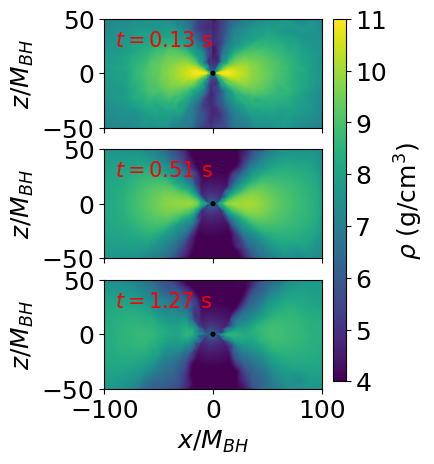}
  \caption{The density of the disk for three different times, 0.13s (top), 0.51s (middle), and 1.27s (bottom) showing the disk drain with time.}
\label{fig:densityvt}
\end{figure}

Over time, the density drops, causing the accretion rate to drop as the disk drains. 
The density of the disk for three different times is shown in Figure \ref{fig:densityvt}. The electron fraction in the disk and the outflow is set by the relative time scale of fluid motion relative to the time scale on which weak processes are occurring. Following \citet{Miller2020}, we compute the weak time scale as 
\begin{equation}
  \label{eq:def:tau:+:-}
  \tau_\pm = \frac{\rho Y_{e}}{G_{Y_{e}}^\pm}
\end{equation}
and the time scale for fluid motion as 
\begin{equation}
  \label{eq:def:tau:a}
  \tau_a(r) =\frac{1}{t_f - t_i}\int_{t_i}^{t_f}dt \frac{\theta_d}{\left\langle v_a\right\rangle_{\rho,Y_{\rm e},\theta,\phi}}
\end{equation}
for characteristic disk opening angle
\begin{equation}
  \label{eq:def:thd}
  \theta_d(t,r) = \sqrt{\frac{ \int_{S^2} \sqrt{-g} d^2x \rho \theta^2}{\int_{S^2} \sqrt{-g} d^2x \rho}}
\end{equation}
and mass-averaged lepton advection velocity 
\begin{equation}
  \label{eq:def:adv:velocity}
  \left\langle v_a\right\rangle_{\rho,Y_{\rm e},\theta,\phi}(t,r) = \frac{\int_{S^2} \sqrt{-g} d^2x \rho Y_{\rm e} u^2}{\int_{S^2} \sqrt{-g} d^2x \rho Y_{\rm e}},
\end{equation}
where $\rho$ is the fluid density, $Y_e$ is the electron fraction, and $G_{Y_e}^+$ and $G_{Y_e}^-$ is the fluid-neutrino interaction rate for weak processes that increase and decrease electron fraction respectively. 
The times $t_f$ and $t_i$ bound the time-average used to compute $\tau$, $\theta$ is the angle off the equator so that $\theta=0$ is the equator and $\theta=\pi/2$ is the north pole. 
$\sqrt{-g}$ is the square root of the determinant of the spacetime metric, and $u^2$ is the theta-component of four-velocity of the fluid. 
See \citet{Miller2019a} for a more detailed description of $G_{Y_e}$ and \citet{Miller2020} for more details on this time-scale analysis procedure. 

\begin{figure}[t]
  \centering
  \includegraphics[width=\columnwidth]{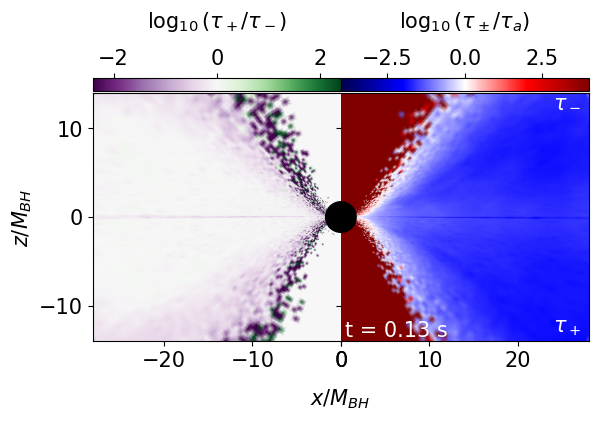}\\
  \includegraphics[width=\columnwidth]{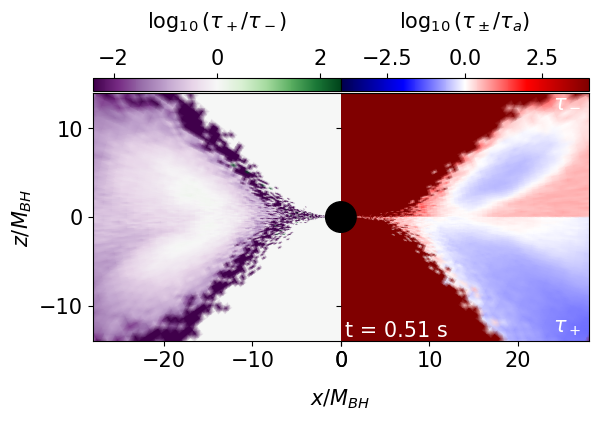}\\
  \includegraphics[width=\columnwidth]{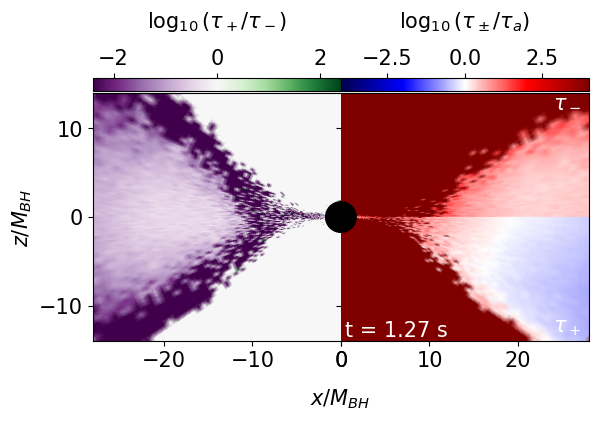}
  \caption{The electron fraction increasing ($\tau_{+}$) and decreasing ($\tau_{-}$) timescales relative to the fluid advection timescale ($\tau_{a}$) at the same three snapshots in the simulation as in Fig.~\ref{fig:densityvt}. }
\label{fig:tau:over:time}
\end{figure}

The top right pane of each panel in Figure \ref{fig:tau:over:time} shows the ratio of $\tau^+$ to $\tau_a$, the bottom right the ratio of $\tau_-$ to $\tau_a$ and the left panel shows the ratio of $\tau_+$ to $\tau_-$. 
In the right pane, a small ratio implies that weak processes dominate. 
As the ratio grows, weak processes become less important in setting $Y_e$ compared to fluid motion, and the electron fraction freezes out. 
The top row shows the disk at 0.13 s, the middle at 0.51 s, and the bottom at 1.27 s. 
The left pane shows that $\tau_+$ is smaller than $\tau_-$, indicating weak processes are driving the electron fraction up. 
However, as the disk cools these weak processes become inefficient compared to fluid motion and the electron fraction in the disk freezes out. 

\begin{figure}[t]
  \centering
  \includegraphics[width=\columnwidth]{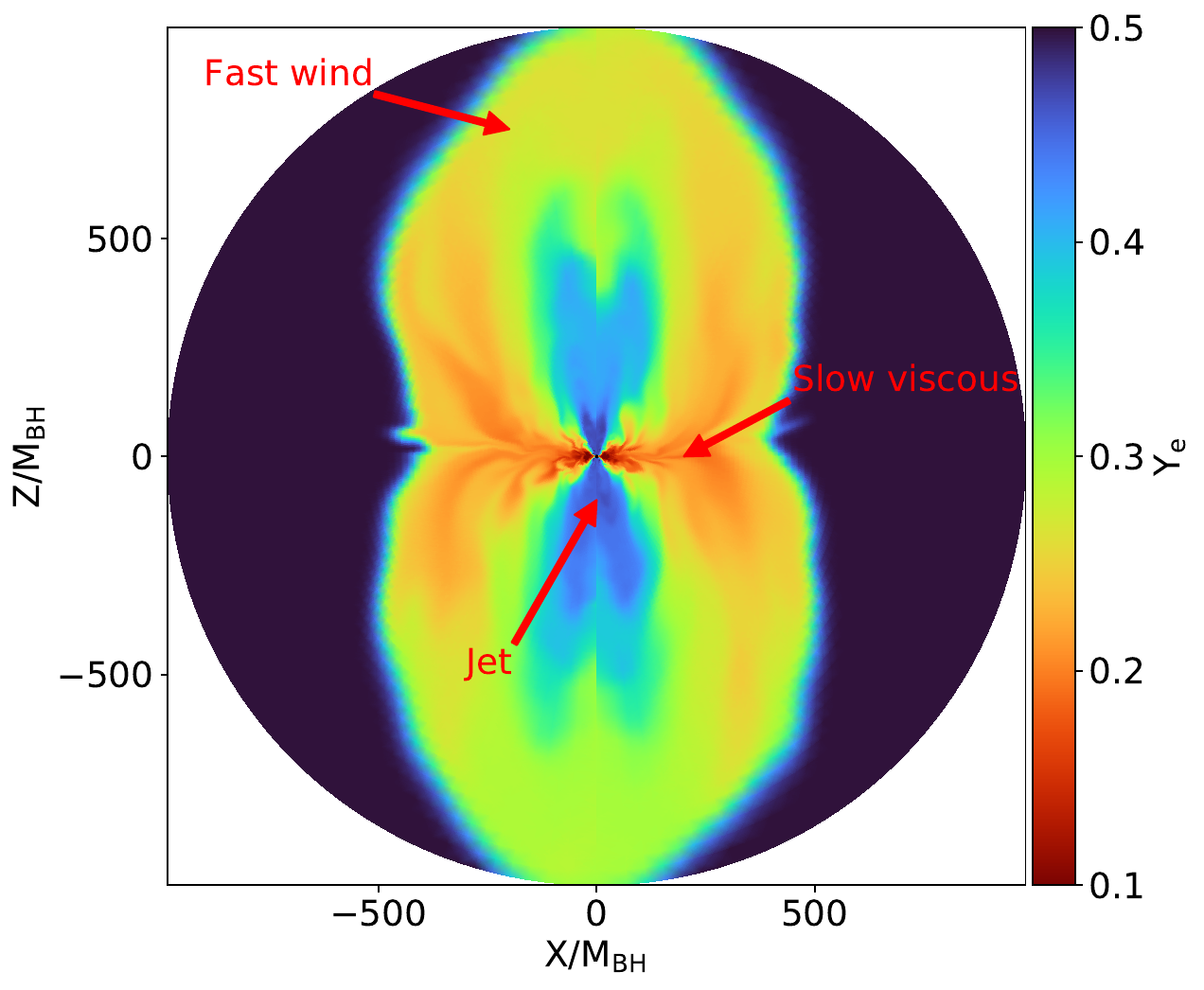}
  \caption{Schematic of the outflow components of the disk. For illustrative purposes, this figure uses a zoomed-out snapshot of the electron fraction $Y_e$ of the disk at t=30\ ms and a contrasting colormap.}
\label{fig:outflow:components}
\end{figure}

Outflows begin to be launched early in the lifetime of the disk, although they travel at different speeds, and thus become gravitationally unbound at different times. 
Figure \ref{fig:outflow:components} sketches these different components out: The magnetic field powers a jet via the \citet{BlandfordZnajek} mechanism; turbulent heating drives a hot, fast disk wind in an hourglass shape out the poles of the disk; and turbulent viscosity drives a slower moving equatorial outflow.  
The viscous mechanism eventually unbinds the most mass. 
In contrast, the jet is the fastest mechanism but unbinds the least mass. 
While we describe these three outflow mechanisms as separate here, in reality these mechanisms are difficult to disentangle and thus uniquely quantify. 

\subsection{From tracer to trajectory}
Once the simulation has completed, we down select tracers which are unbound to study the nucleosynthesis. 
This filter involves the calculations of two physical constraints. 
The first is that the tracer be 250 gravitational radii ($G M_{BH}/c^2$) away from the central black hole. 
The second is that the Bernoulli parameter be $B_e > 0$. 
The Bernoulli parameter originates in modeling of hydrostatic flows. $B_e = 0$ implies hydrostatic equilibrium, $B_e < 0$ implies a flow in-falling into a gravitational potential, and $B_e > 0$ implies a gravitationally unbound flow \citep{Narayan1995}. 

This selection criterion results in 79,556 `short' tracers at 0.12 seconds and 461,690 `long' tracers at 1.2 seconds. 
The difference between these two subsets comes only from running the simulation for an extended duration. 
Over the course of this additional second of simulation time, the ejected mass increases by a factor of 18.5 with the electron fraction decreasing by 0.1 on average. 
The temperature and density also show sizable changes in favor of the production of heavy elements. 
A summary of the difference between short and long evolutions is provided in Table \ref{tab:diffs}. 

\begin{deluxetable*}{ccccccccc}
\tabletypesize{\footnotesize}
\tablecolumns{9}
\tablewidth{0pt}
\tablecaption{ A summary of the ending values for short and long tracer information. Averages are indicated with a \textsuperscript{\textdagger} and are computed by weighting via the mass of each tracer respectively. At the end of a tracer a homologous expansion is employed. \label{tab:diffs}}
\tablehead{
\colhead{Name of run} & \colhead{Stop time} & \colhead{Number of traces} & \colhead{Total Mass} & \colhead{Tracer Mass\textsuperscript{\textdagger}}   & \colhead{Ye\textsuperscript{\textdagger}} & \colhead{Entropy\textsuperscript{\textdagger}} & \colhead{$T_9$\textsuperscript{\textdagger}} & \colhead{$\rho$\textsuperscript{\textdagger}} \\
\vspace{-0.8cm}
 &\colhead{(s)} & & \colhead{(g)} & \colhead{(g)} & & \colhead{($k_B$/baryon)} & \colhead{(GK)} &  \colhead{(g/cm$^3$)}
}
\startdata
Short & 0.12 & 79556                     & $3.557 \times 10^{30}$ & $4.471 \times 10^{25}$ & 0.247 & 19.67 & 1.76 & $2.05 \times 10^5$ \\
Long & 1.2 & 461690                    & $6.567 \times 10^{31}$ & $1.422 \times 10^{26}$ & 0.146 & 15.49 & 1.98 & $7.75 \times 10^5$ \\
\enddata
\vspace{-0.8cm}
\end{deluxetable*}

If a tracer is found to be unbound in the short case, it is also unbound in the long case (by the definition of being unbound using the above two constraints). 
The bulk of the tracers, 382134 = 461690 - 79556, become unbound on timescales greater than 0.1 seconds, owing to the dynamics of the central engine. 
Magnetohydrodynamics (MHD) disk models typically drive an early, fast outflow powered by heat and magnetic forces \citep{Siegel2017,Christie2019} and a late, slow outflow powered by turbulent viscosity \citep{ShakuraSunyaev1973}. 
The latter outflow is enhanced by nuclear recombination, incorporated into the NSE finite temperature equation of state \citep{Fernandez2019,Fahlman2022,Just2022,Haddadi2023}. 
Our disk is no exception, and the more massive late-time outflow is from the slower viscous mechanism.

The total amount of mass unbound in the `short' tracers is significantly smaller than in the `long.' 
At 0.12 s, when the `short' tracers are extracted, the disk has accreted roughly $9.57\times 10^{31}$ g of mass. The mass in the `short' tracers accounts for about 3.7\% of that accreted mass. 
At 1.2 s, when the `long' tracers are extracted, the disk has accreted $9.81\times 10^{31}$ g of mass, only a small fraction more (this is due to the power law decay shown in Figure \ref{fig:mdot}). 
However, the total mass in the `long' tracers is $6.46\times 10^{31}$ g, or 65\% of the accreted mass and about 27\% of the total mass of the disk. 
Other late-time models, such as \citet{SiegelMetzger2018, Fernandez2019, Christie2019}, indicate late-time outflow can be as much as 40\% of the disk mass. 
Our result, as well as the other literature, indicate that extrapolating total mass in the outflow at late times based on early-time mass flux will introduce inaccuracies of about an order of magnitude. 
This is likely due to the different velocities of the outflow, as the fast-moving outflow is less massive than the slower-moving outflow. 

We take the set of traces and convert them into a `trajectory' for use in post-processing nucleosynthesis. 
A trajectory extends the temperature and density profiles contained in each tracer by assuming a homologous expansion. 
The simulation of nucleosynthesis for a given trajectory, however, does not start at the point of homologous expansion. 
Instead, the starting point of our nuclesynthesis calculations begins at the last time the temperature drops below $T = 10$ (GK). 

A homologous expansion is implemented as follows.
The velocity is assumed to be constant, yielding an increment of the Cartesian coordinates after a duration of time, $dt$, $dx_i = v_i \times dt$. 
The density is extrapolated as a power law, $\rho \sim \rho_e/t^3$, where $\rho_e$ is the density at the time of extrapolation. 
The temperature is extrapolated from the density assuming an ideal gas with $\Gamma=5/3$.\footnote{The ideal gas equation of state is \textit{only} used to extrapolate the temperature to produce trajectories.} 
As a consequence of these assumptions, the final time points associated with trajectories are independent from one another and do not interact hydrodynamically (unlike a tracer). 

The end point of tracers (starting point of of the homologous expansion) vary drastically. 
This situation arises naturally from the simulation and thus means the conditions under which heavy element synthesis proceeds will also show large variation. 
Our results thus highlight the need for future nuclear sensitivity studies to cover a wide range of conditions, as shown in the recent work of \citet{Li2022}. 

The additional impact of radioactive heating from nuclear processes can be substantial, and result in a change in the temperature evolution of the trajectory relative to a homologous expansion. 
Nevertheless, it is expected to be a larger effect for dynamical ejecta than in disk ejecta \citep{Lippuner2015}. 
For this reason, it will be considered in subsequent work. 

\subsection{Nuclear inputs}

We use Portable Routines for Integrated nucleoSynthesis Modeling (PRISM) to model $r$-process nucleosynthesis \citep{Sprouse2021}. 
The nuclear input to PRISM is based on the 2012 version of the Finite Range Liquid Droplet Model (FRDM) \cite{Moller2012, Moller2016}. 
Neutron induced reactions, including radiative capture and fission, are calculated with the CoH$_3$ statistical Hauser-Feshbach code \cite{Kawano2019, Kawano2021a, Kawano2021b}. 
Rates of $\beta$-decay, $\beta$-delayed fission, and the associated probabilities to emit neutrons are calculated assuming a statistical de-excitation from excited states \citep{Mumpower2016, Mumpower2018}. 
The REACLIB database is used for secondary reaction rates \citep{Cyburt2010}. 
Conditions suitable for a robust fission recycling $r$-process are not found in this work. 
Therefore, a symmetric 50/50 split is used for fission products in order to increase the computational efficiency of PRISM without impact to the resultant nucleosynthesis nor any of our conclusions. 

\section{Results}

First we describe the differences in key astrophysical quantities that influence the nucleosynthetic outcomes by running for longer times. 
We then analyze the nucleosynthesis itself. 

\begin{figure}[t]
  \centering
  \includegraphics[width=\columnwidth]{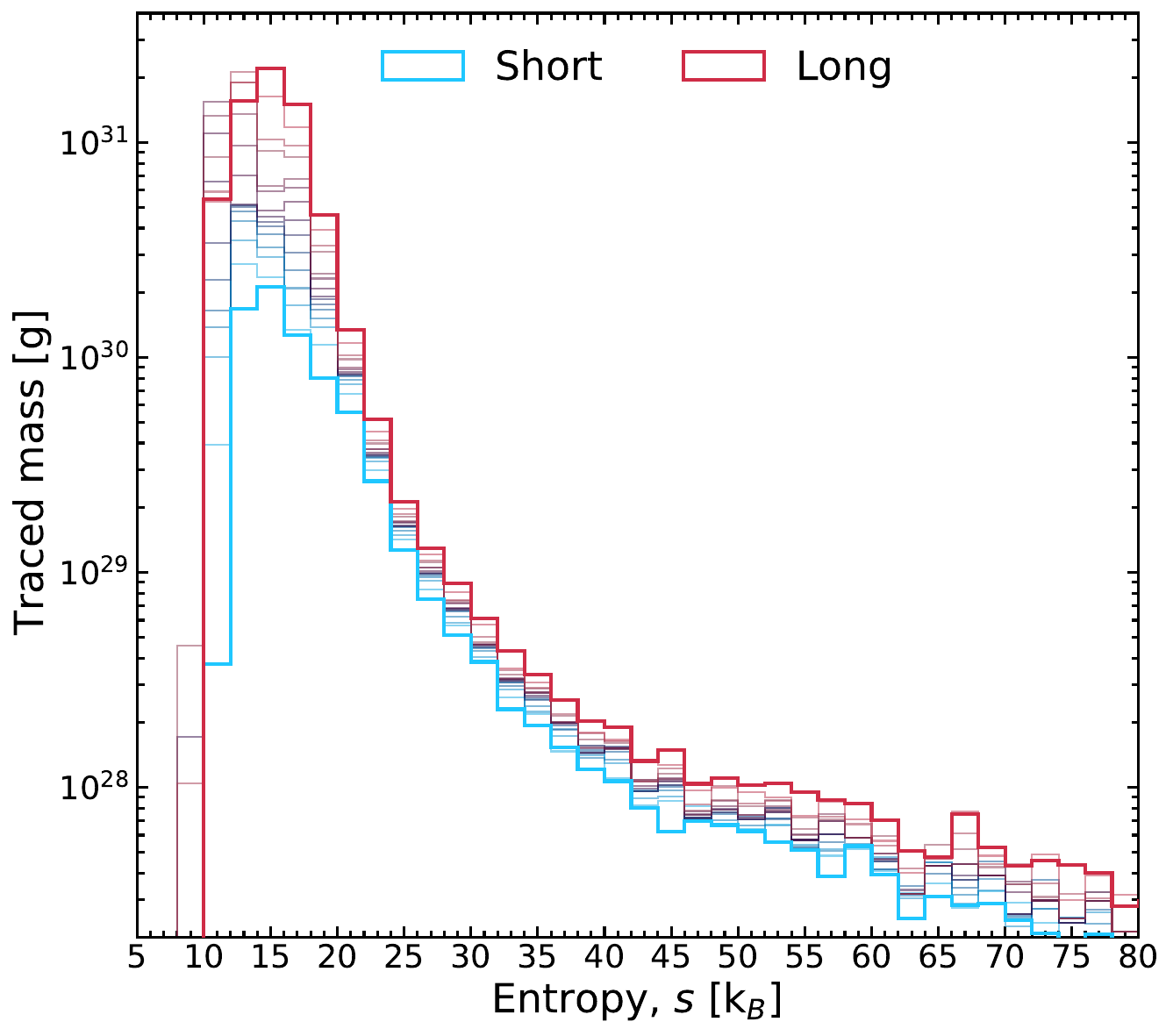}
  \caption{Comparison of entropy distributions of tracers between the short and long runs. The short cut (0.12 s) is shown in blue and long cut (1.2 s) in red. Intermediate snapshots of the entropy distribution are shown between these two snapshots.}
\label{fig:s_dist}
\end{figure}

In Figure \ref{fig:s_dist} we present the difference in the entropy distribution of the unbound tracers between the long and short runs. 
The short-duration run has overall less mass ejected, which can be seen from the lower maximum value on the y-axis for traced mass. 
In addition, the long-duration simulation has a lower average entropy (15.49 $k_B$/baryon as compared to 19.67 $k_B$/baryon in the short case). 
The shift to lower entropy values as the simulation runs longer arises due to the different ejection mechanisms---the early-time, fast outflow is more thermally, as opposed to viscously driven, and may contain a component of material entrained in the jet. 

\begin{figure}[t]
  \centering
  \includegraphics[width=\columnwidth]{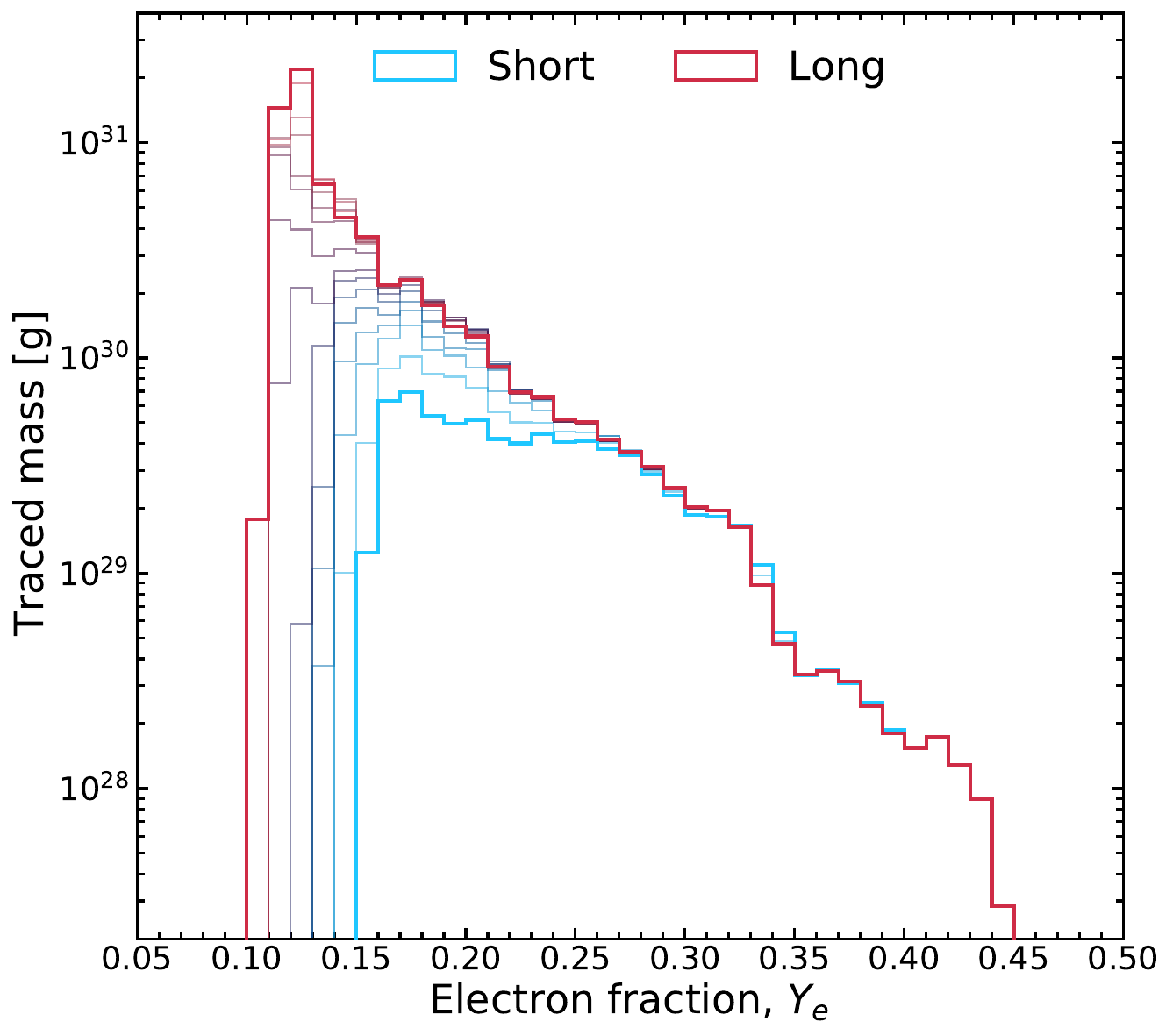}
  \caption{Comparison of the electron fraction distributions for the short cut (0.12 s, blue) and the long cut (1.2 s, red). Intermediate snapshots of the $Y_e$ distribution are shown between these two snapshots.}
\label{fig:ye_dist}
\end{figure}

Also crucial to the resultant nucleosynthesis is the value of the electron fraction at the end of the tracer. 
Figure \ref{fig:ye_dist} compares the distributions of electron fraction between the two cases. 
We find that the longer duration simulation has significantly lower $Y_e$ than the short duration simulation due largely to viscous material which became unbound later in the simulation. 
This strong shift to lower $Y_e$ is a harbinger of subsequent heavy element formation. 

The additional low entropy and low $Y_e$ tracers that are captured in the long-duration run will have slightly different typical nucleosynthetic evolutions as compared with the short tracers. 
Firstly, a lower electron fraction, with all else being equal, means more neutrons available for capture on seed nuclei and a more robust r-process. 
This effect is enhanced by lower entropy which means that material will fall out of equilibrium sooner and experience a more robust r-process. 
Finally, not only does a lower entropy produce a more robust r-process, it also changes the shape and position of the peaks in the distribution \citep{Mumpower2012a, Mumpower2012b, Orford2018, Vassh2020, Vassh2021}. 

We now turn to the assumption of homologous expansion and contrast the results between early and late times. 
The evolution of the temperature and density profiles is critical in the first few seconds as the resultant nucleosynthesis occurs almost entirely in this timescale \citep{Kajino2019, Sprouse2022}. 

\begin{figure}[t]
  \centering
  \includegraphics[width=\columnwidth]{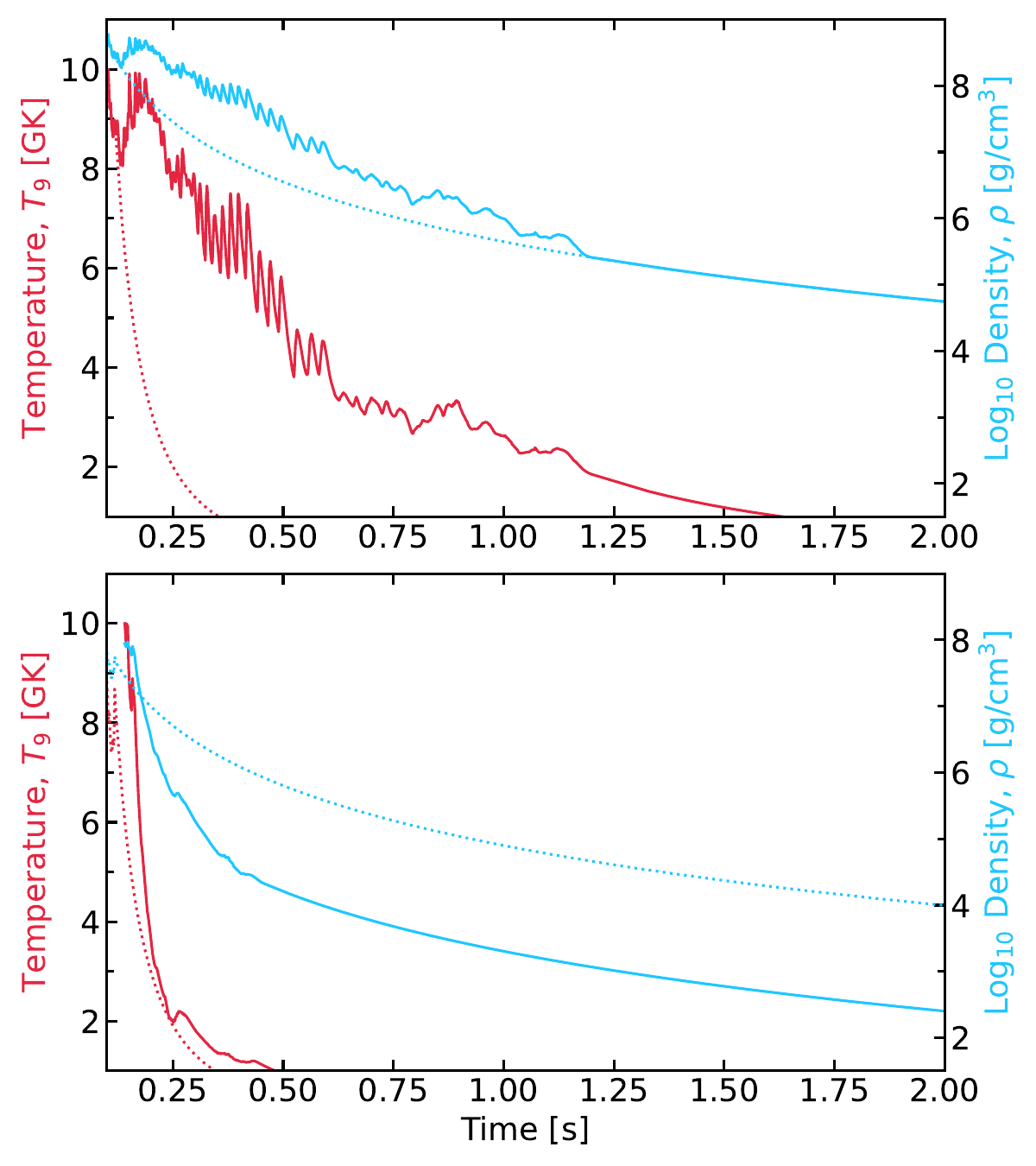}
  \caption{Differences between the homologous expansion assumption for short and long trajectories. The dotted lines indicate the short run while solid lines indicate the long run. The top panel shows a case where both the $T_9$ and $\rho$ evolution is greatly impacted. The bottom panel shows a case where the $\rho$ is greatly impacted.}
\label{fig:traj_diff}
\end{figure}

In Figure \ref{fig:traj_diff} we highlight two individual trajectories. 
The top panel shows a case where the temperature and density evolution are both altered. 
In this panel, the long cut (solid) maintains a higher temperature and density for longer than the short cut (dotted). 
The longer time spent in the 3 GK to 1 GK region means the $r$ process is ``hotter," spending more time in $(n,\gamma) \Longleftrightarrow (\gamma,n)$ equilibrium.
In addition, the long-duration tracer spends more time at higher density but by happenstance, lands on the homologous expansion curve derived from the short tracer. 

The bottom panel of Fig.~\ref{fig:traj_diff} shows a case where the density is orders of magnitude more defuse in the long run as compared to the short, although the temperature drops off similarly as one would expect from homologous expansion using the short cut. 
In this case, due to the drop in density, the long cut nucleosynthesis is less robust than the short duration cut. 

In general we find the longer cuts behave as a combination of the temperature and density profiles shown in the two panels of Figure \ref{fig:traj_diff}. 
On average, the material in the long cut experiences higher densities at later times with a marginally higher temperature evolution as compared with the short cut. 
The final two columns of Table \ref{tab:diffs} highlight these differences where the temperatures are roughly comparable but the density is a factor of 4 larger. 

Nuclear reactions scale as the square of the density \citep{Rauscher2001}, so that reaction rates in the long cut are $\sim 16$ times faster than in the short cut. 
Furthermore, the higher densities are occurring at later times, when reaction rates are more likely to be out of equilibrium, thus substantially favoring more neutron-rich nucleosynthesis \citep{Mumpower2012c}. 
We find the increase in density at late times to be the primary driver of the differences in the nucleosynthetic outcome between the short and long cuts. 

\begin{figure}[t]
  \centering
  \includegraphics[width=\columnwidth]{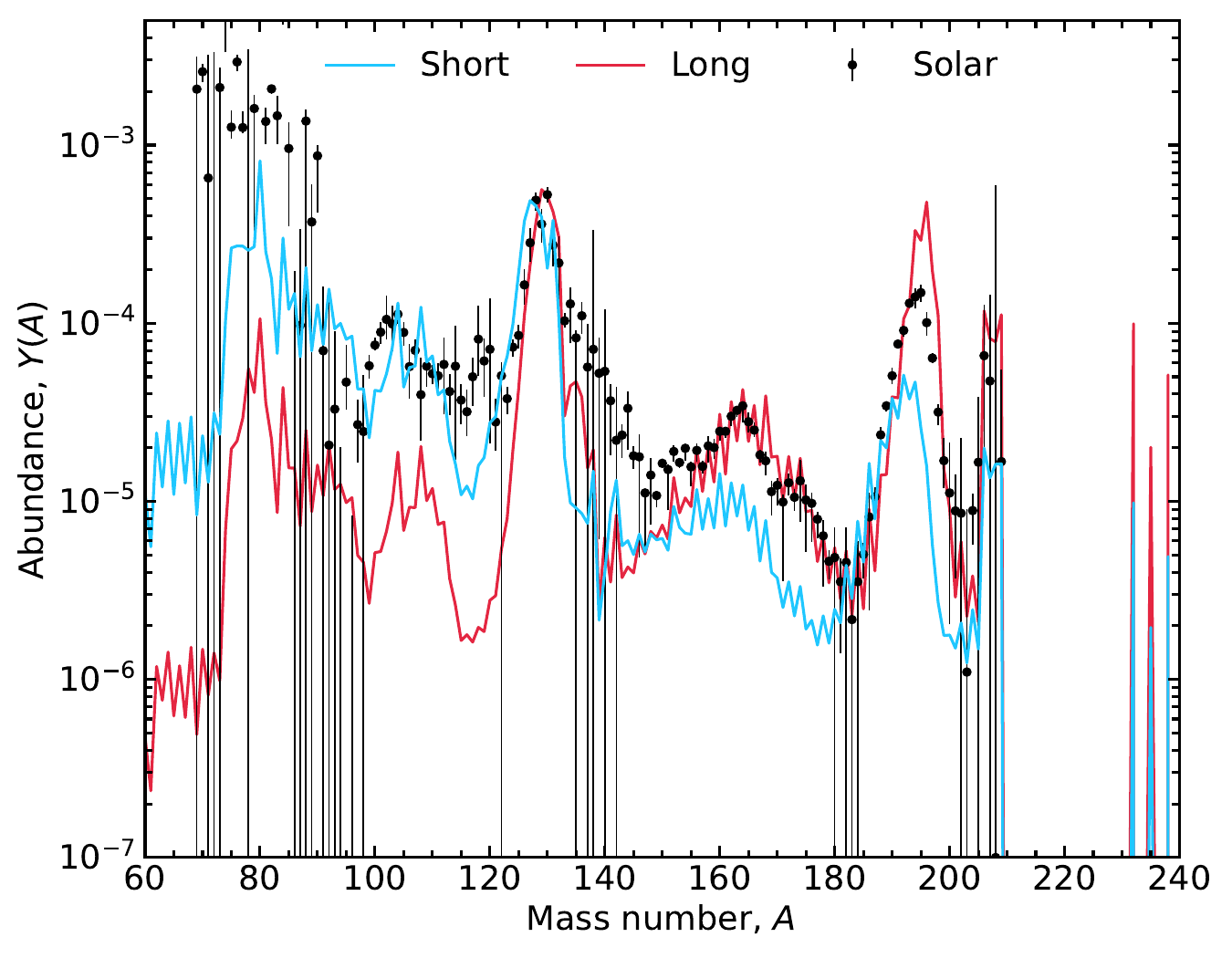}
  \caption{Mean final isotopic abundances at 1 Gyr from the complete ejecta of a NS-BH accretion disk. Solar data in black. }
\label{fig:nucuncert_tot_A}
\end{figure}

\begin{figure}[t]
  \centering
  \includegraphics[width=\columnwidth]{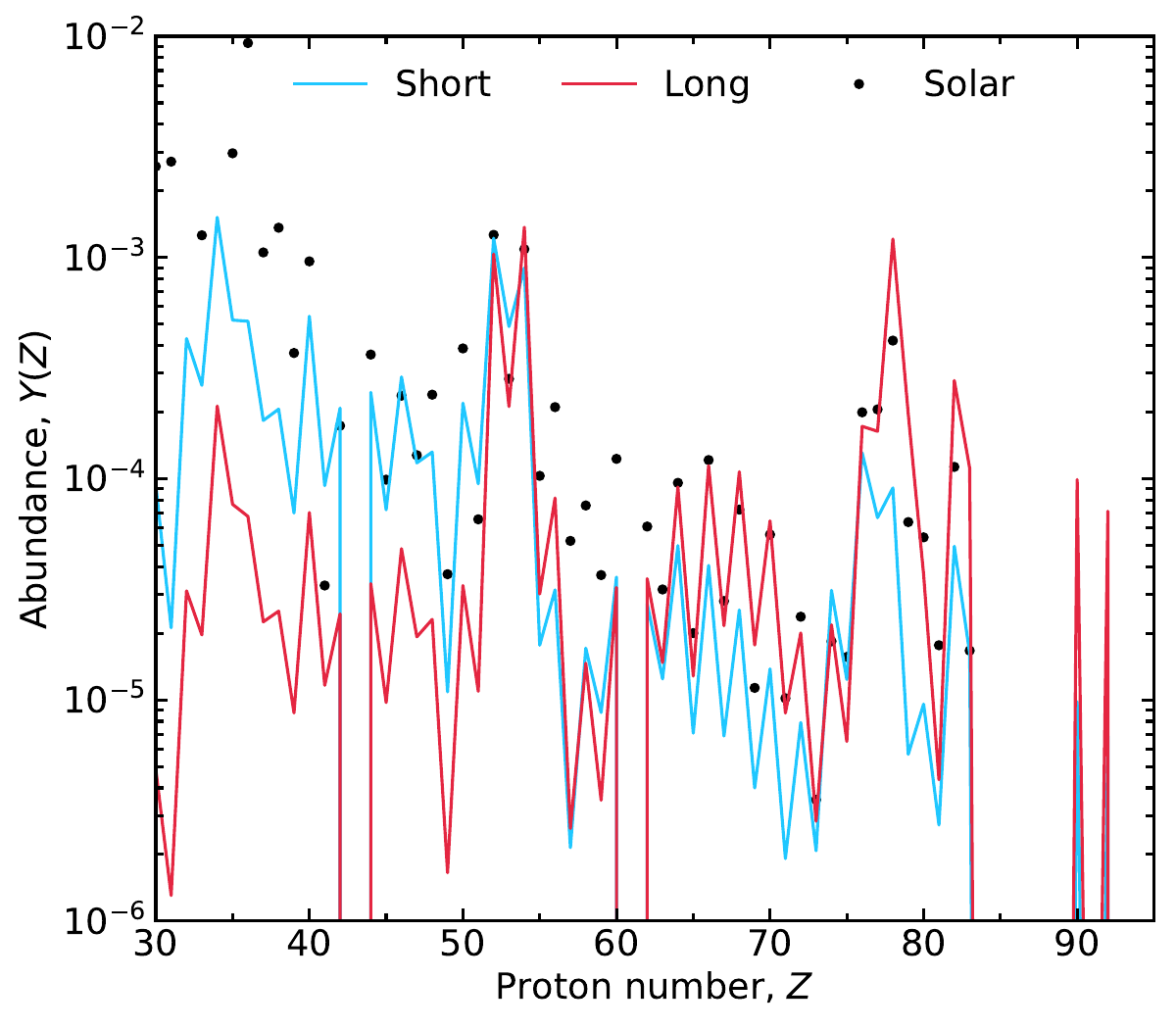}
  \caption{Mean final elemental abundances at 1 Gyr from the complete ejecta of a NS-BH accretion disk. Solar data in black.  }
\label{fig:nucuncert_tot_Z}
\end{figure}

The final abundances for the total mass ejected in each case are shown in Figure \ref{fig:nucuncert_tot_A}. 
The associated elemental abundances are shown in Figure \ref{fig:nucuncert_tot_Z}. 
We find a more robust $r$ process ensues as emergent viscous material emanates from the disk. 
The short scenario has a first peak where elements like strontium reside with a reduced  third peak production. 
In contrast, the long simulation shows a complete $r$ process through the actinides, albeit with a reduced first peak. 
While the actinides are produced in substantial quantity, we do not find evidence of fission recycling. 
Instead, material just makes it to superheavy nuclei ($A\sim280$) which ultimately decay to populate the longer-lived actinides \citep{Holmbeck2023a}. 
In this simulation superheavy elements ($Z>103$) are not found in sufficient quantity  to impact a kilonova signal \citep{Holmbeck2023b}.

The elemental pattern, in particular, shows abundances regions that are clearly simulation-uncertainty-dominated (large variation between short and long cuts). 
This spans nearly the entire pattern from the weak $r$-process peak ($A\sim80$), to the lanthanides, the third peak ($A=195$) and the actinides, while the second peak ($A=130$) remains relatively unaltered. 

There are also points that cannot be readily explained by simulation uncertainties, since the results from the different cuts of the simulation cannot account for the remaining discrepancy from the solar residuals. 
In particular, the lighter elements of a `weak' $r$-process component below $Z=50$ as well as the transition nuclei that reside between the second $r$-process peak and the lighter lanthanides ($50 \lesssim Z \lesssim 60$), have larger errors from nuclear physics uncertainties than seen from the simulation. 
Additionally, nuclear physics models like FRDM2012 have a closed $N=126$ shell far from stability, which in this simulation results in an overproduction of this peak relative to the solar residuals. 
Relevant nuclear physics uncertainties for $r$-process nucleosynthesis have been studied extensively in the works of \citet{Mumpower2016r, Vassh2019, Misch2021, Mumpower2022}. 

\begin{figure}[t]
  \centering
  \includegraphics[width=\columnwidth]{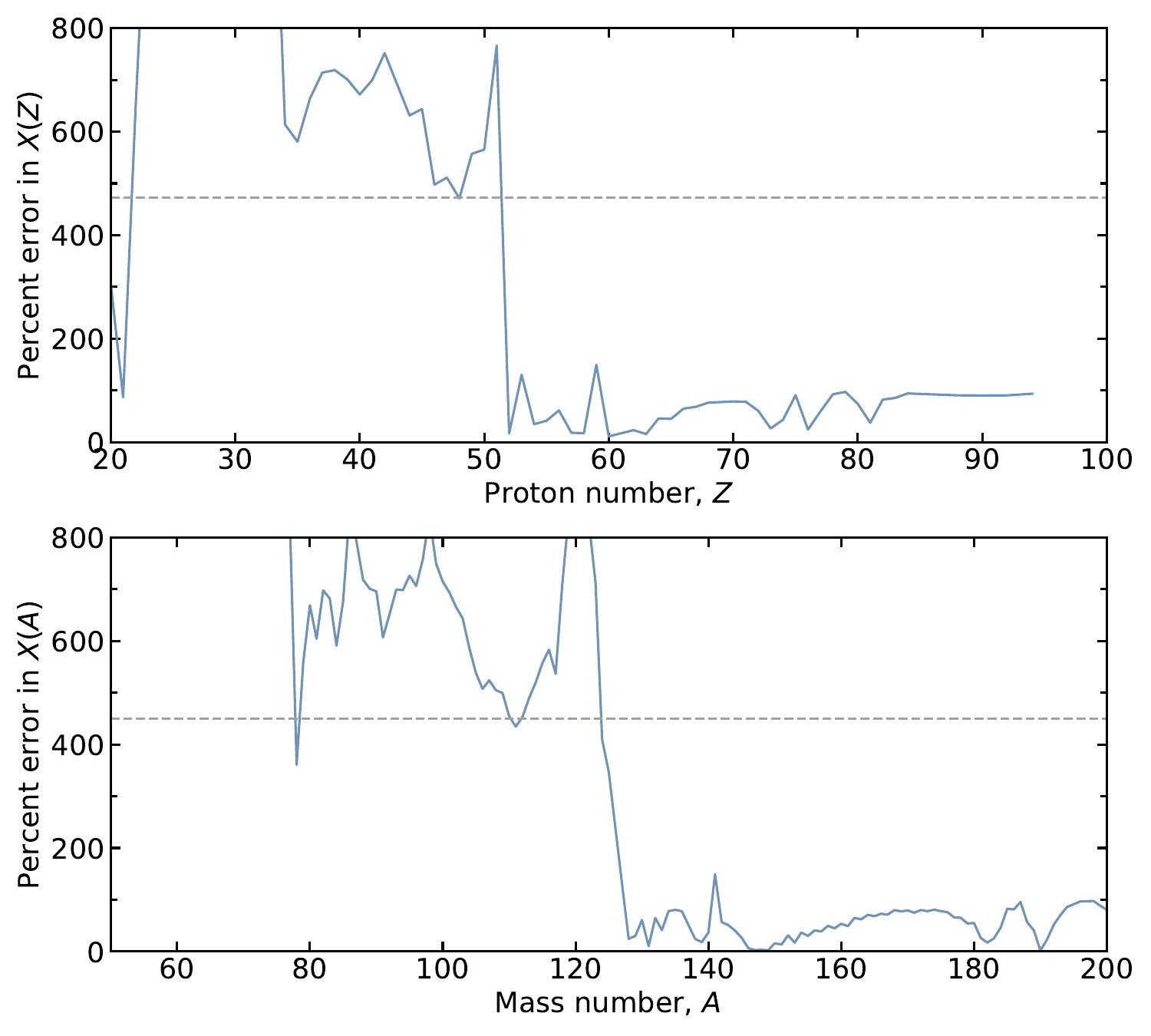}
  \caption{Percent error in the mean final mass fractions as a function of $Z$ or $A$ when using short duration simulation. Average values of these functions are represented by the dashed grey lines. }
\label{fig:pe_za}
\end{figure}

We now quantify the error between the 0.12 second cut and 1.2 second cut by calculating the percent error, $\delta = |\frac{X_{0.12}-X_{1.2}}{X_{1.2}}| \times 100$, where $X_j$ are the respective final mass fractions. 
Figure \ref{fig:pe_za} shows this value as  a function of proton number (top panel) and mass number (bottom panel). 
The average percent error for both functions is between 450 and 500\% as indicated by the dashed grey lines. 
Discrepancies can be found throughout the pattern, but are especially astounding for lighter nuclei. 
A useful rule of thumb derived from this calculation is that for nuclei $Z \leq 50$, the error in population is roughly a factor of 6 while for $Z > 50$, the error in population is roughly a factor of 2. 

\begin{figure}[t]
  \centering
  \includegraphics[width=\columnwidth]{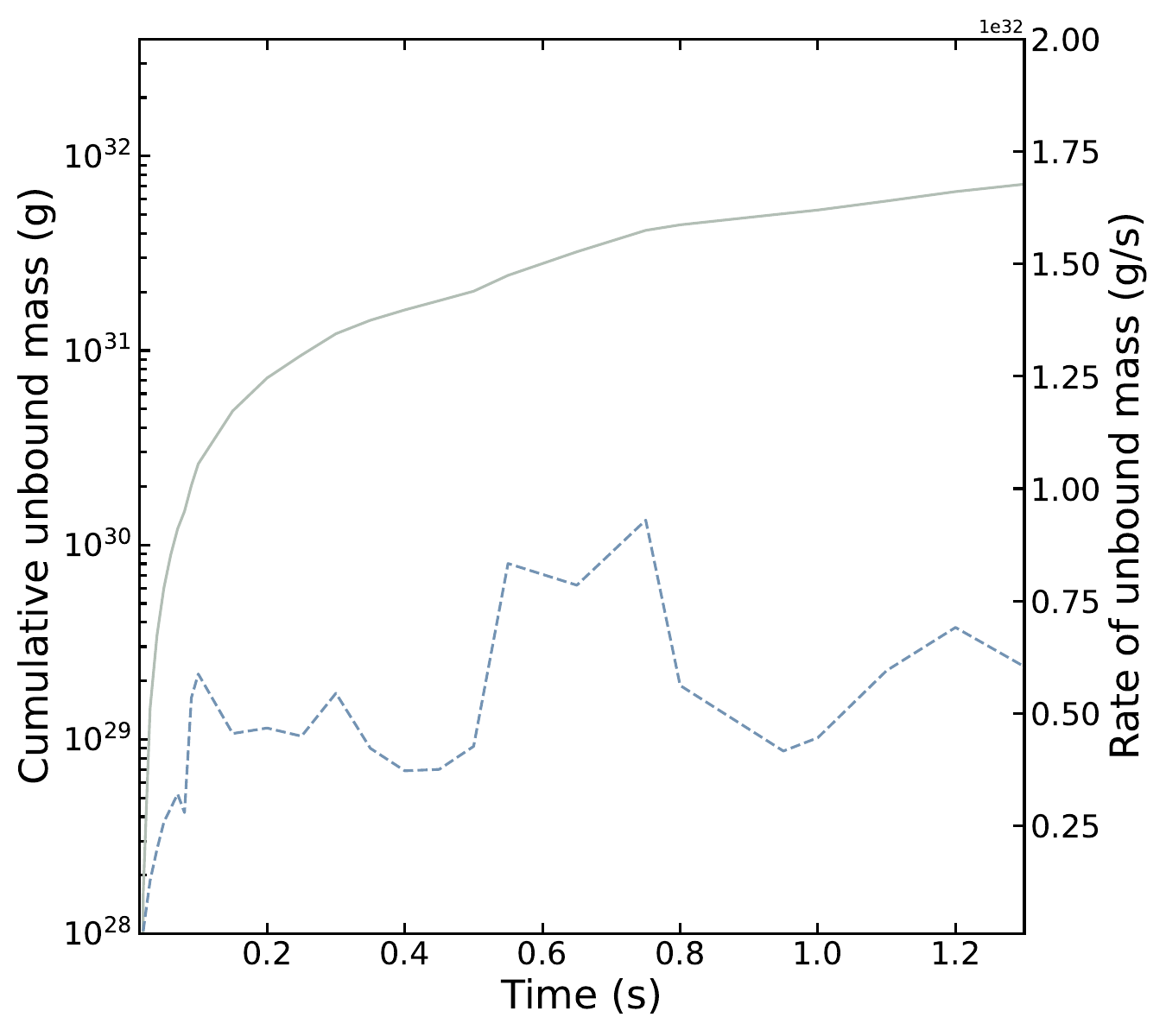}
  \caption{The accumulation of unbound mass (solid grey; left Y-axis) and the rate of unbound mass (dashed blue; right Y-axis). }
\label{fig:cum_mej}
\end{figure}

We now address the question of whether or not the stopping point of our simulation at 1.2 seconds (the long cut) is complete. 
By complete we mean that ejecta has stopped impinging on the extraction surface in sufficient amount such that the electron fraction and other relevant distributions would begin to asymptote, leaving the  nucleosynthesis unchanged. 
To gauge this behavior, we plot in Figure \ref{fig:cum_mej} the cumulative mass ejected (grey curve read from the left Y-axis) as a function of time at the extraction surface.
The derivative of this quantity, or rate of unbound mass ejection (dashed blue curve) is also shown and can be read from the right Y-axis. 
While the cumulative mass ejection looks to be slowing down, it is important to to note that the salient feature of this curve is its log scale. 
The bulk of the unbound material arrives at the extraction surface at later times, and the derivative has yet to approach zero. 
We conclude that simulations must be run for longer times to fully capture the extent of unbound material.

\section{Conclusion}

We have simulated a black hole accretion disk system resulting from a binary neutron star merger for 1.2 seconds using full transport neutrino radiation magnetohydrodynamics (GR$\nu$RMHD). 
We have analyzed the resultant nucleosynthesis, which is greatly impacted as compared with the same simulation cut at 0.12 seconds. 
While we find the total amount of unbound ejecta has yet to completely asymptote (Figure \ref{fig:cum_mej}), our results provide the first insights of running nucleosynthesis with a long duration simulation. 
In particular, we find that emergent viscous material in the plane of the disk to be primarily responsible for the vastly different nucleosynthetic outcome between the short and long duration cuts. 

Our work shows that by running simulations to later times, lanthanides are produced in similar proportion to the first peak (weak) $r$ process. 
To obtain conditions favorable for lighter element production that is inline with the solar pattern, one needs additional processing via neutrinos (that is not found in our simulation), or some other physical mechanism. 
Monte Carlo transport in \nubh{} is only performed in regions of the engine where weak processes are subdominant compared to fluid motion, i.e., when $Y_e$ has frozen out. 
However, on the time scale of the longer simulation (1s), it is possible these slower processes matter, and we may be under-counting them. 
This is one possible source unaccounted for neutrino processing.

We note that late-time lanthanide-rich outflow from this post-merger disk does not change the fact that the fast-moving lanthanide-poor ejecta may produce an early blue component to a kilonova \citep{Miller2019b}. 
Moreover, these results cannot be straightforwardly extended to the collapsar case, where the disk is fed and the thermodynamic conditions vary as a power law with time \citep{Miller2020}.

In the near future longer duration high-fidelity simulations will become common place. 
We have shown that late-time modeling is required to fully capture the richness of phenomenology in the nucleosynthesis and neutrino sector, and we look forward to continual developments in the community to uncover the details regarding the origin of the heavy elements. 



We thank Luke Roberts for valuable feedback on the initial draft of this manuscript. 
This work was supported through the Laboratory Directed Research and Development program under project numbers 20220564ECR and 20230052ER at Los Alamos National Laboratory (LANL). 
LANL is operated by Triad National Security, LLC, for the National Nuclear Security Administration of U.S. Department of Energy (Contract No. 89233218CNA000001). 
GCM and KAL acknowledge support by the Department of Energy Office of Nuclear Physics award DE-FG02-02ER41216 and by the the Fission in r-Process Elements (FIRE) topical collaboration in nuclear theory, funded by the U.S. DOE, contract No. DE-AC5207NA27344.  
GCM acknowledges support by the Network for Neutrinos, Nuclear Astrophysics and Symmetries (N3AS) through the National Science Foundation Physics Frontier Center Grant No. PHY-2020275 and by the U.S. Department of Energy, Office of Science, Office of Advanced Scientific Computing Research and Office of Nuclear Physics, Scientific Discovery through Advanced Computing program under Award Number DE-SC00268442 (ENAF). 
This work was partially supported by the Office of Defense Nuclear Nonproliferation Research \& Development (DNN R\&D), National Nuclear Security Administration, US Department of Energy.
KAL gratefully acknowledges the support of the U.S. Department of Energy through the LANL/LDRD program and the Center for Nonlinear Studies for this work. 
We thank the Institute for Nuclear Theory at the University of Washington for its kind hospitality and stimulating research environment. 
This research was supported in part by the INT's U.S. Department of Energy grant No. DE-FG02- 00ER41132. 
This work is approved for unlimited release with LA-UR-23-28417.



\bibliographystyle{aasjournal}
\bibliography{refs}

\end{document}